\documentclass[12pt]{article}
\usepackage[margin=1in]{geometry}
\usepackage{graphicx}
\usepackage{breqn}

\def\Vec#1{{\bf #1}}
\def\lesim{\ \hbox to 0 pt{\raise .6ex\hbox{$<$}\hss} \lower.5ex\hbox{$\sim$}\ }
\def\gesim{\ \hbox to 0 pt{\raise .6ex\hbox{$>$}\hss} \lower.5ex\hbox{$\sim$}\ }
\def\erfc{{\rm erfc}}

\title{Non-linear Plasma Wake Growth of Electron Holes}

\author{I H Hutchinson, C B Haakonsen, and C Zhou}

\begin{document}
\maketitle

\begin{abstract}
  An object's wake in a plasma with small Debye length that drifts
  \emph{across} the magnetic field is subject to electrostatic
  electron instabilities. Such situations include, for example, the
  moon in the solar wind wake and probes in magnetized laboratory
  plasmas. The instability drive mechanism can equivalently be
  considered drift down the potential-energy gradient or drift up the
  density-gradient.  The gradients arise because the plasma wake has a
  region of depressed density and electrostatic potential into which ions are
  attracted along the field. The non-linear consequences of the
  instability are analysed in this paper. At physical ratios of
  electron to ion mass, neither linear nor quasilinear treatment can
  explain the observation of large-amplitude perturbations that
  disrupt the ion streams well before they become ion-ion unstable. We
  show here, however, that electron holes, once formed, continue to
  grow, driven by the drift mechanism, and if they remain in the wake
  may reach a maximum non-linearly stable size, beyond which their
  uncontrolled growth disrupts the ions.  The hole growth
  calculations provide a quantitative prediction of hole profile and
  size evolution. Hole growth appears to explain the observations of
  recent particle-in-cell simulations.

\end{abstract}

\section{Introduction}

The wake behind an object in a plasma that drifts perpendicular to the
applied magnetic field is filled in by plasma flow along the field
from either side. This flow produces a characteristic multidimensional
potential well structure in the wake that attracts ions and repels
electrons. The approximate steady-state form of supersonic wake
potentials of separated ion streams has been established through
one-dimensional models for
decades\cite{Whang1968,Gurevich1969,Gurevich1975} and more recent work
has established the subsonic solution requiring multiple
dimensions\cite{Hutchinson2008b,Hutchinson2010}. However, there
remains a great deal of uncertainty about the \emph{stability} of the
wake to unsteady short wavelength electrostatic perturbations.  The
solar-wind wake of the
moon\cite{Bosqued1996,Ogilvie1996,Halekas2005,Halekas2011} is a
classic naturally-occuring example of this wake problem, and in-situ
satellite measurements have observed various electrostatic
fluctuations in it\cite{Kellogg1996,Tao2012}. Several large-scale
computational
simulations\cite{Farrell1998,Birch2001,Kallio2005,Kimura2008} have
also shown wake instabilities, but their nature has been
controversial. The purpose of the present work is to provide a
detailed explanation of the mechanisms that drive the wake
instabilities. These may have important applications also for
magnetized laboratory plasmas and their interactions with probes.

The idealized configuration we study is represented by a magnetized
plasma flowing perpendicular to the field but normal to a flat, thin,
object\cite{Hutchinson2012}. This is equivalent to a plasma flowing
with a sufficiently high Mach number past a spherical (or similar
approximately unity aspect ratio) object. The high cross-field
velocity in this second case, causes the object to be thin relative to
the characteristic lengths in the wake.  In other words, the sphere is
strongly compressed in the flow direction, when measured in
appropriately scaled units. The analysis represents the plasma
velocity distribution function in one dimension, along the assumed
uniform magnetic field. The plasma is presumed to drift in the
transverse, wake direction, with simply a uniform drift velocity. So
there is a one to one correspondence between downstream position and
time since passing the object's position. The Debye length is much
smaller than the object.

\begin{figure}[htp]
  \centering
  \includegraphics[width=0.6\textwidth]{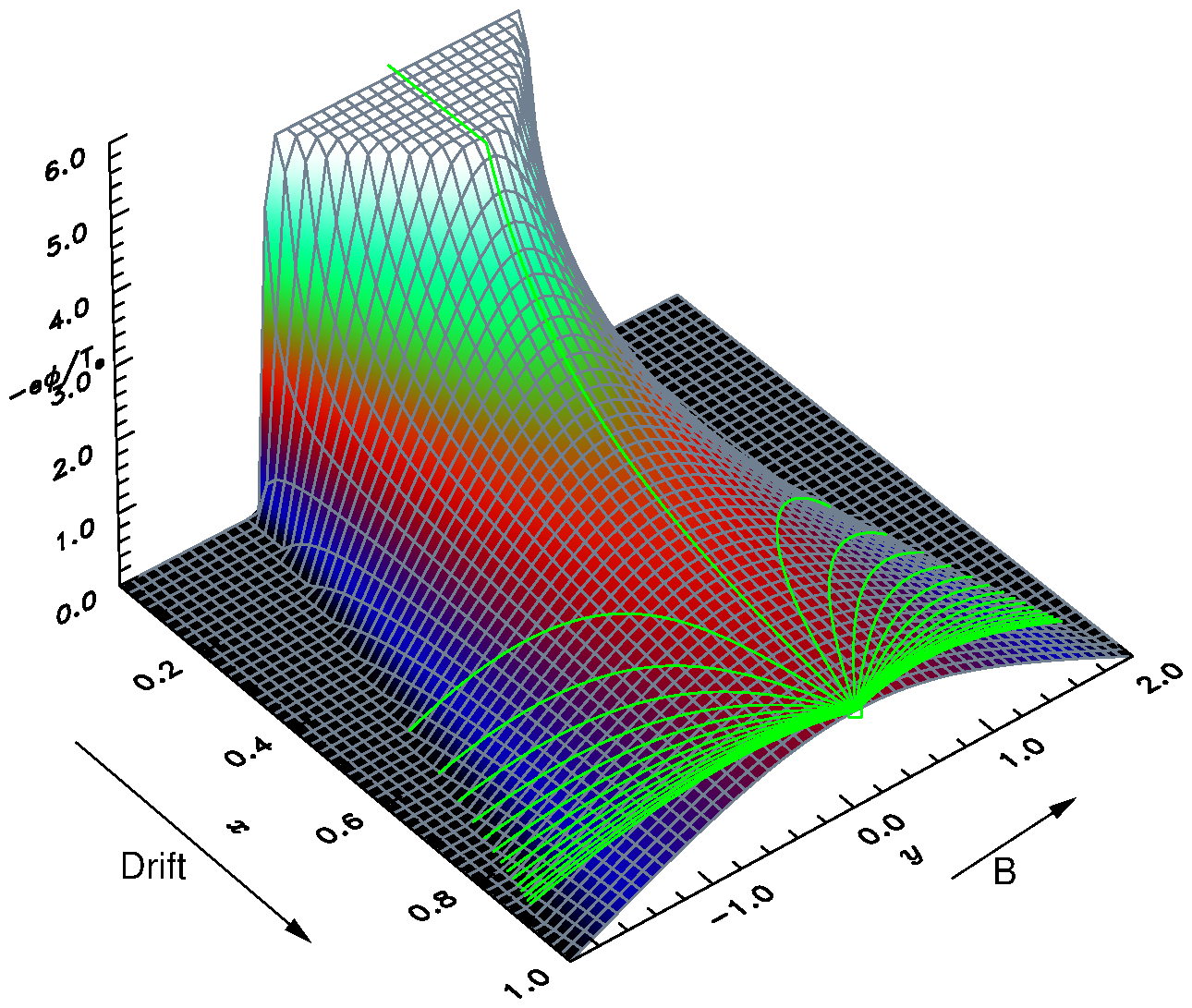}
  \caption{Potential energy structure of electrons as a function of
    ($x$) distance down the wake and ($y$) position along the magnetic
    field. (Potential is cut off for viewing purposes at height 5). Example
    collisionless orbits that arrive at a position on the wake axis
    are shown. Those that have low parallel energy at the final
    postion are substantially de-energized. Artificial electron/ion
    mass ratio of $1/20$ is used to assist the visualization.}
  \label{potlstructure} 
\end{figure}

A self-consistent wake potential develops that attracts ions and
repells electrons illustrated in Fig.\ \ref{potlstructure}.
In a previous paper\cite{Hutchinson2012} it has been shown by
integration along orbits that the collisionless electron distribution
in the wake potential structure acquires a depression that is
localized in velocity. We here call this localized reduction of
$f_e(v)$ the ``dimple''. It is generated on electron orbits that are
near the threshold of being reflected by the potential energy
hill. These orbits climb the hill, converting their parallel kinetic
energy into potential energy. Then, because they approach the peak
with very small parallel velocity (they are nearly or just reflected)
they spend a long time near the ridge of the potential, and during
that time their transverse drift carries them down the potential
ridge. Eventually their parallel motion carries them down off the
ridge, but not before they have substantially reduced their total
energy compared with when they climbed it. They have experienced
``drift de-energization''. By contrast electron orbits that are far
from the threshold of reflection, having either much more or much less
total energy than the potential ridge, spend much less time on the
potential ridge. They are far less de-energized. The distribution
function is constant along orbits in a collisionless plasma. So if the
external distribution is monotonically decreasing in kinetic energy
(e.g. a Maxwellian) then an orbit that started (outside the potential
structure) at a higher total energy (because of de-energization) has a
phase-space density $f(v)$ \emph{lower} than orbits that have
experienced less de-energization. This is the qualitative explanation
of the mechanism forming the dimple. Its form was calculated
quantitatively by numerical orbit integration in reference \cite{Hutchinson2012}.

The distribution function dimple that arises is linearly unstable to
Langmuir waves. Therefore one expects this de-energization effect to
excite electrostatic instabilities, which will have a tendency to fill
in and smooth out the distribution non-linearly until the growth rate
is suppressed. Because the dimple size depends strongly on the
electron to ion mass ratio, the free energy available prior to
non-linear saturation (which was calculated) also depends on mass
ratio; simulations that use artificially low mass ratio are therefore
liable to obtain unphysically large fluctuation levels.
For true mass ratios the energy available to
instabilities is only $10^{-4}$ to $10^{-3}$ of the electron thermal
kinetic energy; so the level of Langmuir
wave turbulence expected is modest.

The purpose of the present work is to pursue further the non-linear
development of the instability driven by this de-energization
mechanism so as to explain what is observed in recent large-scale
simulations of this problem\cite{Haakonsen}. Those simulations clearly
observe the formation of the electron distribution dimple, but the
observations are of course of its self-consistent non-linear
state.
\begin{figure}[htp]
  \centering
  \includegraphics[width=.6\textwidth]{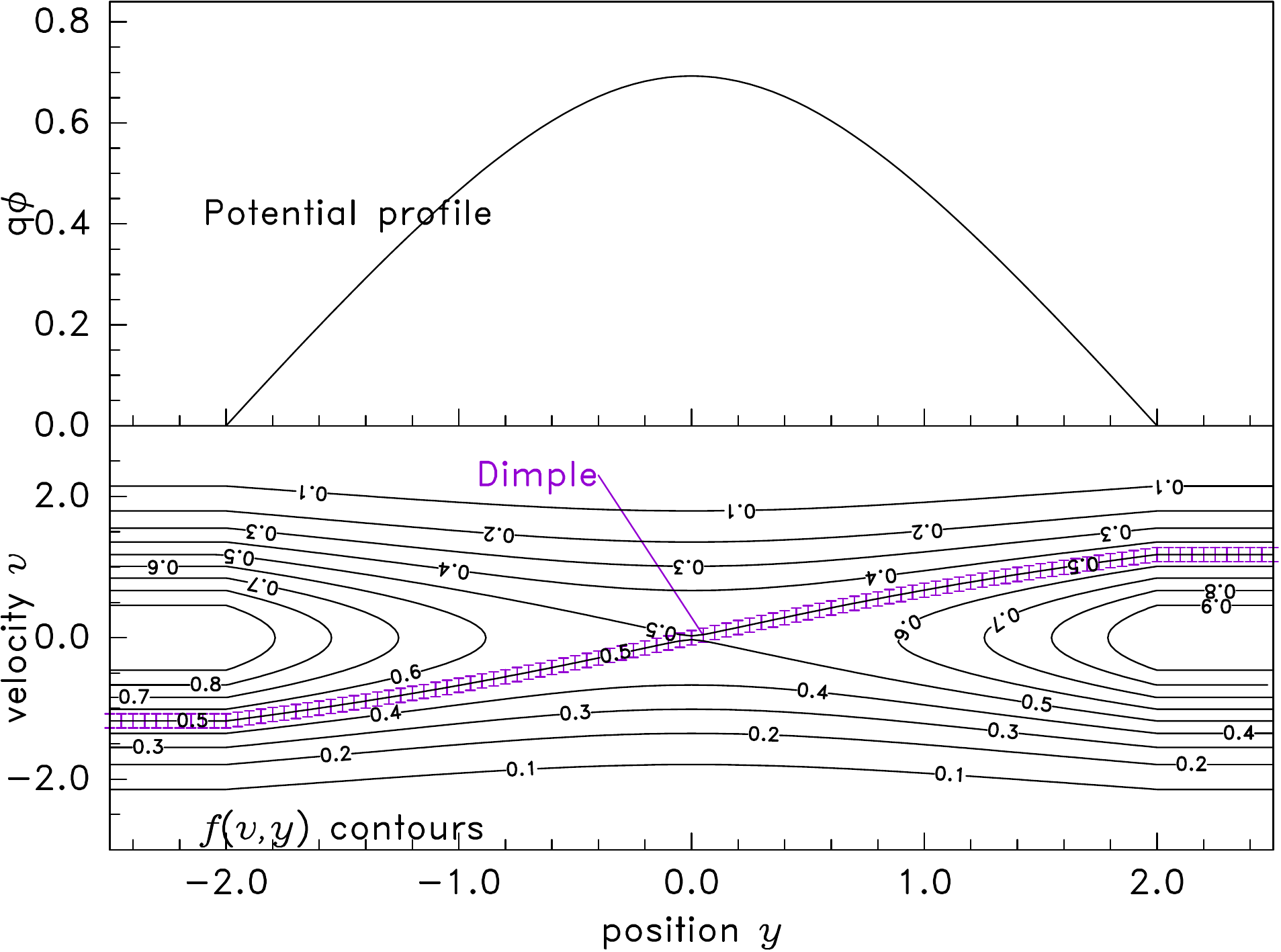}
  \caption{Illustrative electron phase-space orbits in the vicinity of
    the wake's potential energy hill. The hill extends along
    the parallel $y$ coordinate to the edge of the wake. The dimple
    lies along the lower left to upper right branches of the
    separatrix, hatched.}
  \label{fig:hillcontour}
\end{figure}
In phase space (see Fig.\ \ref{fig:hillcontour}), the dimple is
centered along the separatrix contour of constant total energy that
the wake potential defines.  In addition to incoherent noise, we
observe small persistent coherent structures, like eddies, localized
to the dimple in phase space. They propagate in along the
dimple with approximately the local electron velocity (and
acceleration). These structures, which (we will show) are electron
holes, therefore leave the simulation, moving to large parallel
distance, quite quickly. The exception to this behavior is that holes
move much more slowly near the x-point of the energy contours (where
the phase-space velocity is zero) which is naturally at the ridge of
the wake potential structure. As the simulations progress (down the
wake) eventually one (or more) of the electron holes near the x-point
grows to a large size, and disruption of the ion velocity distribution
occurs. There is a large amount of free energy in the ion
distribution, because the ions are in two streams, of modest energy
spread, with opposite velocities ($>c_s$). The strong hole growth and
disruption of these streams occurs at a place where (time when)
linearized calculation indicates that the ion streams are
\emph{stable} because of their large separation. The puzzle that our
current analysis addresses is how the perturbation becomes large
enough to disrupt the ion streams well before they themselves become
linearly unstable. Our answer is that the mechanism is a non-linear
one involving electron holes.

In section \ref{bolzmannform} we formulate the Boltzmann equation for
the parallel electron distribution, and solve it approximately
analytically in a potential of specified shape to find the electron
dimple at the potential ridge. This solution supplements the prior
numerical orbit integrations\cite{Hutchinson2012} by providing an
analytic form for the dimple, in particular its velocity-width. In
view of the substantial approximations required to achieve this
analytic solution, section \ref{parallelint} approaches the problem
instead, by an integration in the parallel direction rather than along
the two-dimensional orbits.  This alternative (and equivalent)
formulation shows that the drift (de-energization) effects can be
conceptualized as a term in the one-dimensional Boltzmann equation of
approximately the ``Krook'' collision form. Solving this equation
gives an identical expression for the dimple, through a conceptually
different set of approximations.

The second formulation is more useful for incorporating the effects of
presumed quasilinear diffusion filling in the dimple. Section
\ref{qlinsect} explains the expected consequences of a self-consistent
level of incoherent turbulence. It is shown that this system \emph{cannot}
explain the growth of perturbations to sufficient amplitude to disrupt
the ions and tap into their energy until a place on the wake is reached
where the ion streams are very close to linear instability. In other
words, it cannot explain what is observed in the simulations. 

Section \ref{holesect} provides an explanation and analysis
of the coherent electron holes, and shows that the drift
de-energization mechanism can be equivalently regarded as the drift
convection of holes into regions of higher background density. This
effect causes holes to grow in depth and velocity width. The
self-consistent growth of hole width with background density and the
resulting hole profile (for quasi-neutral holes) is calculated
analytically for Maxwellian background electrons and beam ions. Holes
that retain their integrity and remain near the ridge of the wake
potential structure (the x-point) \emph{can} grow to sizes sufficient
to disrupt the ion streams when the density increase is of the order
of one e-folding. Moreover their characteristics are consistent with
what is observed in the simulations. Therefore we interpret the
pre-linear-threshold disruption of the ion streams as caused by the
long-term non-linear growth of electron holes until they become
energetic enough to tap the ion free energy.

\section{Solving For the Dimple in Two Dimensions}
\label{bolzmannform}

\subsection{Boltzmann's Equation with drift and quasilinear diffusion}
Including ad hoc quasilinear velocity-space diffusion\cite{Davidson1972} with coefficient
$D$, we can write Boltzmann's equation as
\begin{equation}
\left({\partial \over \partial t}+ \Vec{v}.\nabla +
\Vec{a}.\nabla_v\right)f =\nabla_v(D\nabla_v f).
\end{equation}
which for the one-dimensional distribution, magnetized case with
coordinate $y$ in the magnetic field direction, and $x$ in the
perpendicular direction becomes
\begin{equation}\label{yorbits}
{d\over dt}f = \left({\partial\over \partial t} + v {\partial \over \partial
  y} - {q\over m} {\partial \phi \over \partial y} {\partial \over
  \partial v} \right)f = -v_x {\partial f\over \partial x}  + {\partial \over
  \partial v} \left(D {\partial f\over
  \partial v}\right).
\end{equation}
Here the right hand side can be considered the additional source terms
in the 1-D Boltzmann equation arising respectively from drift
de-energization and quasilinear velocity space diffusion.
We will consider a time-independent situation: ${\partial \over
  \partial t}=0$ and constant drift $v_x$. Velocity $v$ written
without a subscript refers here to $v_y$, the parallel velocity, and
the distribution function is one-dimensional along $y$.

The orbits are the characteristics of the left-hand side. They are the
paths in (parallel) phase space corresponding to constant energy 
\begin{equation}
{\cal E} = {1\over2}m v^2 + q \phi =const.
\end{equation}
They are most easily found as the contours of constant energy in phase space.

\subsection{Collisionless Orbits in Specified Potential}\label{normalization}

We take parameters to be normalized so that velocities are in units of
the cold ion sound speed $c_s=\sqrt{T_{e}/m_i}$, and potential is in
units $T_e/e$. The perpendicular distance is scaled such that
$x=$distance$/M_\perp$ where $M_\perp=v_x/c_s$. Then the normalized
energy equation for electrons becomes ${\cal E}={1\over 2} m_r v^2 -
\phi$, where $m_r\equiv m_e/m_i$.

The dimensionless potential form is considered to be controlled by
dynamics separate from what happens to the instabilities. The specific
form illustrated in Fig. \ref{potlstructure} is based on an
approximate solution in the form of two expansions of plasma into a
vacuum, patched at the symmetry axis as discussed
previously\cite{Hutchinson2012}. However, the only features of this
potential shape that substantially matter in the present context are
that $-\phi$ is symmetric and single-peaked in $y$, having a known
curvature near the ridge at $y=0$ and being zero beyond a certain
$y$-distance; and that it decays from a large value at $x\sim 0$
monotonically in the perpendicular. i.e.\ downstream wake ($x$)
direction.  So we will simply specify that

$\bullet$ $\phi_0(x)$ is the monotonic
potential at $y=0$;

$\bullet$ the curvature is given in terms of a scale length $w$ by
${\partial^2 \phi\over\partial y^2}=-\phi_0/w^2(x)$;

$\bullet$ and $y=Y(x)$ is the edge of the perturbed potential.

\noindent [A model potential that fits simulation results with
$T_i=T_e$ has $\phi_0=-1/1.2x$, $Y=1+2x$, and $w=(1+2x)/1.66$.]

Now since in normalized units
\begin{equation}
v={1\over c_s} {dy\over dt} = {d y\over dx},
\end{equation}
the equation of the electron orbits in 2-D space may be written in the
vicinity of the ridge, in terms of an expansion  as
\begin{equation}
m_r {d^2 y\over dx^2} ={\partial \phi\over \partial y} = -{\phi_0\over w^2}y
\end{equation}

Because of the smallness of $m_r$ the orbits do not have a large
duration ($x$-extent). So it makes sense to approximate ${\phi_0/
  w^2}$ as a constant, ignoring its $x$-dependence. Then the orbit
can be solved trivially as
\begin{equation}\label{yofx}
y = {v_0\over k} \sinh(k[x-x_0])
\end{equation}
where 
\begin{equation}
k^2 = -\phi_0/m_rw^2
\end{equation}
and $x_0$ and $v_0$ are the position and velocity of the orbit when
$y=0$.  The approximations leading to this expression are not well
justified near the edge of the perturbed region (and not at all
outside it), nevertheless most of the orbit of electrons that cross the
ridge slowly is spent near the ridge. It is therefore reasonable to
use eq.\ (\ref{yofx}) to estimate the orbit duration $X=x_0-x$
(considered the duration either in time or in space, since $x$ and $t$ are
interchangable) from the edge of the perturbed region to $y=0$. It can
then be considered the solution of
\begin{equation}\label{sinhkX}
\sinh(kX) = kY/|v_0|.
\end{equation}
In order for the approximations adopted to
be consistent, both sides of this equation must be large compared with
unity. Therefore $\sinh \approx {1\over2}\exp$ and 
\begin{equation}\label{kX}
kX \approx \ln\left[2kY\over|v_0| \right]= \ln\left[ 2\sqrt{|\phi_0|}\, Y\over\sqrt{m_r}w |v_0| \right].
\end{equation}

\subsection{Resulting Distribution Without Diffusion}

The electron distribution function at $y=0$, when $D=0$, can be deduced by
considering the change in total parallel energy ${\cal E}$ arising
from the perpendicular drift term. To the extent that $y\ll x$
applies to the relevant parts of the orbit (which we've already
assumed to be a good approximation), the potential energy change
arising from cross-field drift (which is the drift de-energization)
can be estimated from the $x$-gradient of the potential \emph{at the ridge}
\begin{equation}
\delta{\cal E}=-\int {\partial\phi\over \partial x} {d x\over dt} dt \approx
-\int_x^{x_0} {d\phi_0\over dx} dx =\phi_0(x_0-X)-\phi_0(x_0) \approx
-{d\phi_0\over dx} X
\end{equation}
(to first order in $X/x_0$). Substituting for $X$ we get
\begin{equation}
\delta{\cal E}\approx -{d\phi_0\over k dx} \ln(2kY/|v_0|)=-{d\phi_0\over
  dx} {\sqrt{m_r}w\over \sqrt{|\phi_0|}}\ln\left(2\sqrt{|\phi_0|}\,
  Y\over\sqrt{m_r} w |v_0|\right).
\end{equation}
This energy is not included in the parallel energy conservation along
the orbit. In other words, denoting the kinetic energy at the start of
the orbit by ${\cal K}_\infty = {1\over2} m_rv_\infty^2$, and when it
reaches the ridge ${\cal K}_0 = {1\over2} m_rv_0^2$, we have
\begin{equation}
{\cal K}_\infty = {\cal K}_0  -\phi_0+\delta{\cal E}.
\end{equation}
Consequently at $x=x_0$, $y=0$, the distribution function $f_0$ is
different from the (presumed) Maxwellian at the orbit start: 
\begin{dmath}\label{dimpleshape}
f_0 / {n_\infty\over \sqrt{2\pi T}} = \exp(-{\cal K}_\infty)
= \exp(\phi_0-{\cal K}_0 -\delta{\cal E})\\
\approx\exp(\phi_0)\exp(-{1\over2}m_rv_0^2)\left[\sqrt{m_r} w |v_0|\over
2\sqrt{|\phi_0|} Y\right]^P ,
\end{dmath}
where
\begin{equation}
  \label{eq:P}
  P= {d\phi_0\over kdx} = {d\phi_0\over dx} {w\over \sqrt{|\phi_0|}} \sqrt{m_r}
\end{equation}
At the characteristic distance down the wake $x_0\sim 1$, all
quantities $\phi_0$, $d\phi_0/dx$, $w$, and $Y$ are of order unity. So
$P$ is of order the square root of the mass ratio $\sqrt{m_r}$.

\begin{figure}[htp]
  \centering
  \includegraphics[width=0.5\textwidth]{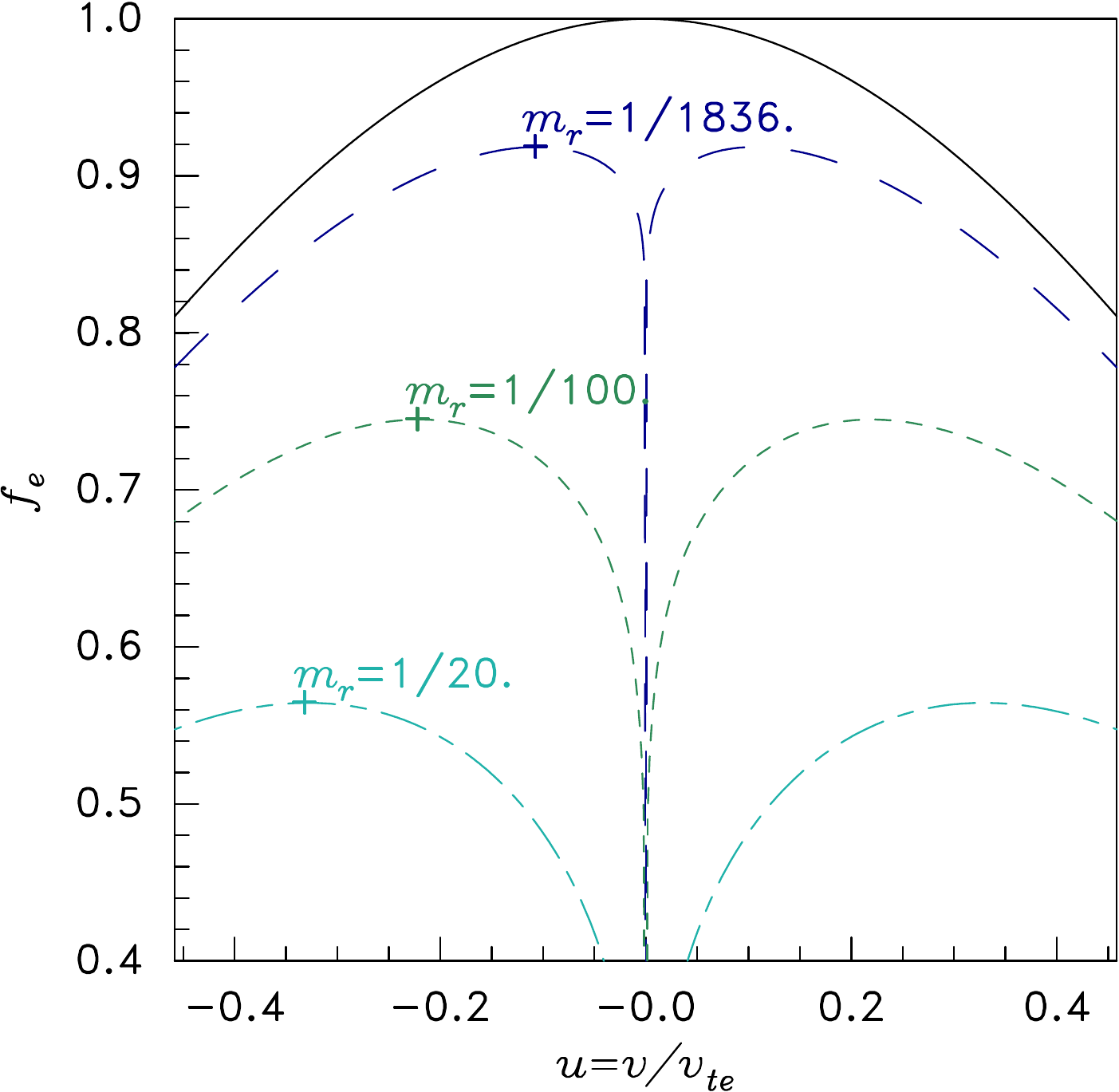}
  \caption{The dimple in the electron distribution at $y=0$, $f_0$ as
    a function of normalized electron velocity $u$, for three
    different electron to ion mass ratios $m_r$. The cross marks the
    dimple width, where $f_0'=0$. The parameters are $\phi_0=-1.32$,
    $d\phi_0/dx=0.96$, $w=1.15$ (so $P=0.96\sqrt{m_r}$), and $Y=2$.}
  \label{dimpleplot}
\end{figure}
The dimple in the electron velocity distribution function is
illustrated in Fig.\ \ref{dimpleplot}, in which electron velocity has
been expressed normalized to its thermal velocity
$u=v_e/\sqrt{2T_e/m_e}= v_0\sqrt{m_r/2}$. The dimple arises in this
Vlasov case as a final multiplication factor on the otherwise
Maxwell-Boltzmann behavior. It actually dominates the behavior near
$v_0=0$ forming a cusp.  Strictly speaking our approximations are
quantitatively unjustified as $u\to 0$, but the qualitative 
observation that a cusp forms is correct.
The approximations also assume
$m_rv_0^2(=2u^2)\ll 1$ and so the factor in brackets is small. But the
power $P$  to
which the factor is raised is also\textsc{\textsc{}} small. The
derivative of $f_0$ with respect to $v_0$ from this expression can
readily be shown to be zero where
\begin{equation}
u^2={1\over 2}m_rv_0^2 = {1\over2}P
\end{equation}
which is a small quantity, of order the square root of the
electron/ion mass ratio, $m_r$. 
This condition may also be written
\begin{equation}\label{velocitywidth}
v_0 =  \left(- {d\phi_0\over dx} {w\over \sqrt{|\phi_0|}}\right)^{1/2} {1\over m_r^{1/4}},
\end{equation}
which indicates the approximate width of the dimple in velocity space.
This width is dictated by the electron/ion mass ratio. Therefore
simulations that use artificially increased mass ratio ($m_r=m_e/m_i\gg
1/1836$) will increasingly misrepresent the electron behavior.

\section{Solving for \protect{$f$} by Parallel Integration}
\label{parallelint}

Thus far we have approached the problem accounting fully for two 
space dimensions and discussing integration along orbits in 2-D space
plus 1-D velocity. Solving the problem analytically has required
major approximations, but has given a reasonable estimate of the
result when there is no velocity-space diffusion.

A different approach to solving for the dimple is to do integration
not along the 2-D spatial orbits but along only 1-D ($y$) in space.
That is actually how eq.\ (\ref{yorbits}) is organized. 2-D orbit
integration takes the first term on the RHS to be part of the orbit
characteristics (and so far has not included the second diffusive
term). By contrast 1-D integration of the equation leaves the
$v_x$-convective term on the RHS, and integrates along a fixed $x$
path in which $y$ and $v$ vary. This is then a truly 1-D (but
phase-space) treatment, but instead of $f$ being constant on orbits,
it varies in accordance with the terms remaining on the RHS.

Within this perspective, we can regard the two terms on the RHS as
being dimple-generating convective de-energization, and quasilinear
diffusion. To some degree they will balance one another: one tending
to form the dimple, the other to smooth it away. The solution for $f$
at some position $y$ can be found in principle by starting in the
unperturbed background region $y_Y$ (actually at $y=Y$, the edge of the
potential perturbation), and integrating orbits inward to
position $y$. In principle the solution is simply
\begin{equation}
f(y,v_y)- f(y_Y,v_Y) = \int -v_x {\partial f\over \partial x}  + {\partial \over
  \partial v} \left(D {\partial f\over
  \partial v}\right) dt.
\end{equation}
This integration must be taken along phase-space orbits of constant
${\cal E}={1\over2}m v^2 + q \phi(y)$, and in practice needs to
be done in terms of position.
\begin{equation}
f(y,v_y)- f(y_Y,v_Y) = \int_{y_Y}^{y} -v_x {\partial f\over \partial x}  + {\partial \over
  \partial v} \left(D {\partial f\over
  \partial v}\right) {dy\over v}.
\end{equation}
This integral determines the difference between the actual $f$ and
the Maxwell-Boltzmann approximation (a Maxwellian scaled by
$\exp(\phi)$). 

A solution by this technique requires us to know what the value of the
terms in the RHS integral are. Focussing first on the convective
de-energization term $v_x\partial f/\partial x$, we don't know its
value exactly until we have the solution everywhere. However, it may
in some circumstances be reasonable to approximate it in a manner
which avoids us having to solve the full-scale integro-differential
system. One such approximation is to presume that the shape of the
dimple changes only slowly with $x$-position. If so, then the dominant
contribution to $\partial f/\partial x$ can be estimated to be the
variation of the overall level of $f$, which is approximately the
Maxwell-Boltzmann. Its variation with $x$ (in normalized parameters)
is $f\propto \exp(\phi)$, in which case
\begin{equation}\label{krookexp}
{v_x\over f}{\partial f\over \partial x} = -v_x{\partial \phi\over \partial
  x} \equiv -\nu_x.
\end{equation}
Here $\nu_x$ is like a collision frequency. And indeed, the term in
the Boltzmann equation to which this approximation corresponds is a
``Krook'' collisional term $-\nu_x f$. Since the orbits of interest
spend most of their time near the potential ridge at $y=0$, it is
reasonable to take $\nu_x$ to be uniform, given by taking
$\phi=\phi_0$. That is the main approximation of this treatment.

Incorporating just this term (i.e.\ taking $D=0$) for now, we can
perform the integral along constant-$x$ based upon the resulting equation
\begin{equation}\label{onedBolt}
{1\over f} \left(v {\partial \over \partial
  y} - {1\over m_r} {\partial \phi \over \partial y} {\partial \over
  \partial v} \right)f=
{1\over f} {d f\over dt} = -\nu_x,
\end{equation}
whose solution is
\begin{equation}\label{fratio}
f(y,v_y)/f(y_Y,v_Y) = \exp[ -\nu_x(t-t_Y)]
\end{equation}
Thus the deviation from Maxwell-Boltzmann distribution can be
considered to be an exponential multiplicative factor whose argument
is proportional to the time an
orbit takes to reach the position $y$.

Notice that the dimensionless form for $d/dt$ in eq.\ (\ref{onedBolt})
shows the terms on the LHS are usually very large compared with the
convective de-energization term (the $\nu_x$ term). That term is
important only where the LHS terms are nearly zero, i.e.\ near $y=0$
where ${\partial \phi\over\partial y}= 0$ and at values of $v$ nearly
equal to zero. In other words, the dimple generation takes place
predominantly at the axis, for velocities near zero there. However, it
is not that the contribution to $df/dt$ is larger there, it is that
the orbit spends far more time there than anywhere else. Passing that
region in phase space contributes most strongly to $t-t_Y$.

The duration, $X=t-t_Y$, of the orbit to the position $y=0$ has
already been solved for in section \ref{normalization}. It was there
taken as an \emph{approximation} that variation of $\partial
\phi/\partial y$ with $x$ could be ignored. Here it is no
approximation, because we are integrating along $x=$constant. Instead the approximation has been made in $df/dx$. In
any case,  we
can immediately appropriate the solution eqs. (\ref{sinhkX}) and
(\ref{kX}) as 
\begin{equation}
X \approx {1\over k}\ln\left[2kY\over|v_0| \right]= \sqrt{m_r\over \phi_0}w \ln\left[ 2\sqrt{|\phi_0|}\, Y\over\sqrt{m_r}w |v_0| \right].
\end{equation}
It should be no surprise that substituting this result into
eq.\ (\ref{fratio}) gives exactly the same dimple as previously:
eq.\ (\ref{dimpleshape}). 

What we've demonstrated, therefore, is that the dimple formation can
be calculated by explicit integration at fixed $x$, based upon an
approximation of the drift de-energization term in a Krook form. This
demonstration gives some additional confidence in the prior treatment. But
it also gives us a more direct way to incorporate quasilinear
diffusion, and (later) to understand electron hole growth.

\section{Quasilinear Electron Velocity Diffusion}
\label{qlinsect}

Now we consider the effects of instabilities that will arise as the
dimple forms and prevent it from ever becoming the deep cusp that the
Vlasov treatment finds. One possible result of such instabilities, if
they consist of many incoherent modes, is an effective quasilinear
velocity-space diffusion. That's the case we discuss first.

\subsection{Self-consistent Diffusion level}

Under this assumption, the physics of the steady state is that the
diffusion magnitude, $D$, adjusts itself corresponding to a moderate
time-independent level of turbulence sufficient to maintain the distribution
function at an approximately neutral stability. It is reasonable (if
the Debye length is small compared to other lengths in the problem) to
assume that the growth rate of the electron instabilities is
intrinsically large compared with other timescales in the problem. In
that case, $D$ must adjust itself so that the instability threshold is
never significantly exceeded. In other words, marginal stability is
always approximately satisfied. 

Solving for the distribution function in those circumstances requires
us to suppose that as we integrate along the orbit we encounter levels
of quasilinear diffusion that are just sufficient to maintain the
distribution function marginally stable.  Doing so requires that in
phase-space regions where the RHS terms are important (i.e.\ mostly
near $y=0$ and $v=0$) the diffusion term counterbalances the
convective de-energization term. 

When the two RHS terms exactly balance
\begin{equation}\label{constraint}
-\nu_x f + {\partial \over \partial v} \left(D {\partial f\over
  \partial v}\right) =0.
\end{equation}
When $D$ is (approximately) independent of $v$, and $f$ deviates only a
small amount from constant, i.e.\ in the vicinity of a shallow dimple,
the solution is of this equation is a parabola. However, we don't
require that the terms exactly balance for orbits whose duration is
sufficiently short that the perturbation introduced by $\nu_x$ is
small.\footnote{One way to model that fact is to allow the product of
  the RHS times the orbit duration to be no larger than some
  appropriate quantity. For example, if we require no more than a
  modest fractional reduction $P$ of $f$ in the dimple, then we must
  take
\begin{equation}
X(v)\left[\nu_x f - {\partial \over \partial v} \left(D {\partial f\over
  \partial v}\right)\right] \le P f.
\end{equation}
Where $X$ is large, the $Pf$ term is small and we recover the previous
condition. But for larger $v_0$, when $P/X$ becomes comparable with
$\nu_x$, the $Pf$ term decouples the diffusion term from any necessity
to balance the $\nu_x$ term, and it can subside to zero. All this
seems rather more elaborate than justified by the current precision.}
The dimple can therefore be considered to be constrained to have a
positive second derivative (equal to $\nu_x/D$) over a region around
$v=0$ that extends to a speed ($|v|$) at which $\nu_x X$ becomes
smaller than of order unity. Outside that velocity region, the second
derivative of $f$ can become negative, as it must in order to merge the
dimple with the bulk of the electron distribution function. The
details of that outer region depend on how quickly the constraint eq.\
(\ref{constraint}) is relaxed, whether $D$ varies with $v$ and so
on. Such details cannot be precisely calculated using the analytic
principles on which this treatment is based. Some ansatz must be
adopted. A simple and plausible one is to choose to represent the
dimple as a negative Gaussian perturbation to the bulk Maxwellian
distribution. The velocity \emph{width} of the dimple Gaussian,
expressed as $2v_d$ such that the Gaussian is $\propto
\exp(-v^2/v_d^2$), is determined by $\nu_xX\sim 1$.

The plasma fluctuation level adjusts the diffusion coefficient $D$ to
achieve marginal stability. If $D$ is small, the dimple is deep,
because its magnitude is such as to give second derivative $\nu_x/D$
near its peak. If $D$ is large the dimple is shallow. Thus, orbit
duration determines the \emph{width}, and marginal stability determines
the \emph{depth} of the dimple. The dimple width is given (see
eq. \ref{velocitywidth}) by $v_d\sim
v_0 \sim m_r^{-1/4}$ (in units of $c_s$) or $u_d\sim \sqrt{2}\;m_r^{1/4}$ (in
units of $v_{te}$).

\subsection{Electron Marginal Stability}

\begin{figure}[htp]
  \centering
  \includegraphics[width=0.45\textwidth]{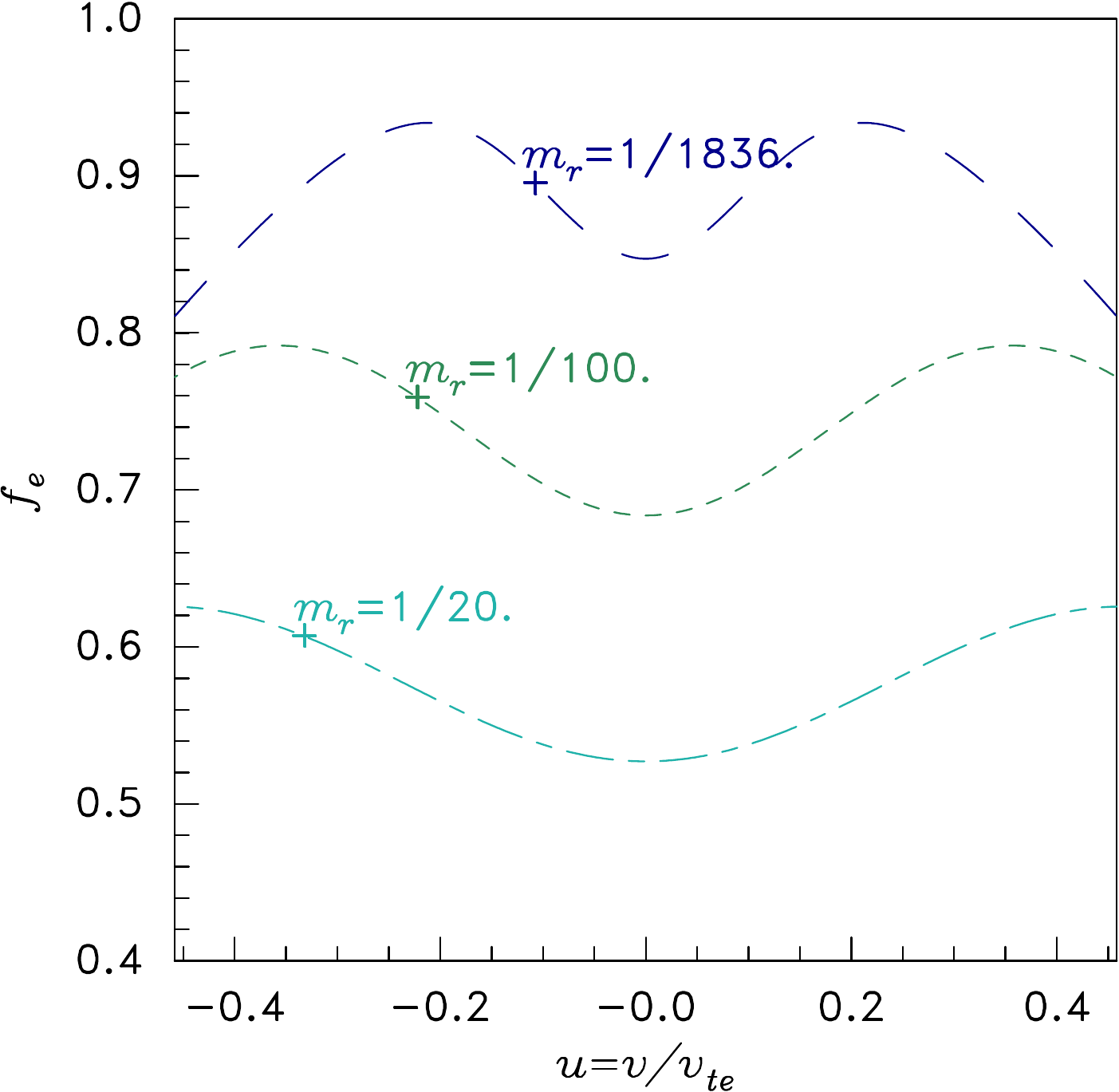}
  \caption{Electron distributions representing a dimple of negative
    Maxwellian form with temperature equal to $m_r^{1/2}T_e$ that are
    marginally stable with immobile ions.}
  \label{marginalfig}
\end{figure}

The dispersion relation of electrostatic waves is
$\epsilon=1+\chi=0$. Instability requires the real part of the
susceptibility $\chi_r$ to be negative at frequency $\omega$ in the
upper half of the complex plane where the imaginary part of $\chi$ is
zero. A bulk Maxwellian electron distribution (ignoring ions for now),
contributes a susceptibility  real-part approximately
$\Re(\chi_{e})=1/k^2\lambda_{De}^2$ (at wave phase velocities small
compared with the electron thermal speed). The contribution from a
dimple Gaussian of temperature $T_d$ and negative density $-n_d$ is
the same but multiplied by $-n_dT_e/n_eT_d$. So the total (real part)
electron susceptibility is
\begin{equation}
  \label{elecsuscept}
  \Re(\chi_{e})= {1\over k^2\lambda_{De}^2} \left[1- {n_d\over n_e}
    {T_e\over T_d}\right].
\end{equation}
Because $k$ is essentially a free choice, it can be adjusted for any
\emph{negative} value of $\Re(\chi_{e})$ to make
$\Re(\chi_{e})=-1$. For a symmetric distribution such as we are
considering, the imaginary part of the susceptibility,
$\Im(\chi_{i})$, is zero at $\omega=0$; so negative $\Re(\chi_{e})$ is
sufficient (as well as necessary) for instability.  Marginal stability
of electrons alone is therefore at $n_dT_e/n_eT_d=1$, which means
\begin{equation}
  \label{marginalgauss}
  {n_d\over n_e} = {T_d\over T_e} \sim m_r^{1/2}.
\end{equation}
However, if we just focus on the zero-velocity peak,
\begin{equation}
  \label{dimplepeak}
  {f_d(0)\over f_e(0)} = {n_d\over n_e}\left(T_e\over T_d\right)^{1/2} 
= \left(T_d\over T_e\right)^{1/2} \sim m_r^{1/4}.
\end{equation}
The dimple depth at marginal stability should be a quite noticeable
decrease in the distribution function at zero velocity, fractionally
$1/1836^{1/4}\sim 1/7$. Fig.\ \ref{marginalfig} illustrates some
cases. Comparison with Fig.\ \ref{dimpleplot} shows that these depths, 
if anything, somewhat overestimate what is expected from filling in the
collisionless dimple by diffusion.

A crude upper bound on the magnitude of the potential perturbation
that will produce quasilinear diffusion sufficient to maintain
marginal stability can be estimated as follows. Quasilinear diffusion
presumes the cumulative effect of stochastic orbits produced by
multiple modes of different phase velocities. How many modes are
involved is uncertain, but what is certain is that it is at least
greater than one, and that therefore the amplitude of any one unstable
mode is insufficient of itself to flatten the distribution (at
marginal stability). A mode whose phase-space island size is equal to
the width of the dimple ($\sim2v_d$) \emph{is} large enough on its own
to flatten the distribution. The single-mode perturbation sufficient
to create such an island width is (in normalized units)
$\tilde\phi\sim m_r^{1/2}$ and is the upper bound of the perturbed
potential at quasi-linear marginal stability. This potential makes a
very small perturbation to the ions, a fractional energy perturbation
of only $m_r^{1/2}/v_i< m_r^{1/2}\sim 1/50$. 

More generally, quasilinear velocity-space diffusivity of particles by
a resonant spectrum of waves of specified electric field (or
potential) is proportional to the inverse square of the particle mass. The
ion streams' velocities place them inside the dimple, subject to the
same resonant spectrum of waves as the electrons. They will experience
a diffusivity smaller by a factor $m_r^2=1836^{-2}$. Negligible
ion perturbation occurs at electron marginal quasilinear turbulence
levels. The free energy of the ions cannot be tapped by quasilinear
electron instabilities.

\subsection{Ion susceptibility contribution}

When the ions have effectively a two-stream distribution in the region
under consideration, they contribute further to instability by
negative contribution to the real part of the susceptibility. The
contribution for low-temperature equal-density beams of velocity $\pm
v_i$ (in units of $c_s$) is 
\begin{equation}
  \label{ion-suscept}
  \Re(\chi_{i}) \approx - {1\over k^2\lambda_{De}^2} {1\over v_i^2} .
\end{equation}
This is sufficient to make a system with purely Maxwellian electrons
unstable when $v_i\le 1$, which is the upper edge of the ion-ion
instability region in a ``Stringer'' plot\cite{Stringer1964}\footnote{Incidentally, the
  electron-ion instability which slightly overlaps the ion-ion
  instability on a standard Stringer plot is \protect\emph{suppressed}
  by the flattening or hollowness of the electron distribution}. If the ion speed is not a
great deal higher than this threshold, then the ion contribution
modifies the electron marginal stability condition, rendering it:
\begin{equation}
  \label{ionelecstab}
    {n_d\over n_e} = {T_d\over T_e}(1-1/v_i^2)  \sim m_r^{1/2}(1-1/v_i^2).
\end{equation}
The marginal-stability depth of the dimple is somewhat decreased. And
to make the dimple shallower in the presence of constant $\nu_x$, the
magnitude of of the quasilinear diffusion coefficient, $D$, must be
larger by the factor $v_i^2/(v_i^2-1)$.

Nevertheless, when $v_i$ (the ion mach number) substantially exceeds 1
(the upper limit for ion-ion instability), the ion susceptibility
contribution does not change the linear marginal stability condition
by very much. It does not much enhance the required quasilinear
diffusivity nor the level of turbulence required to produce it,
and it does not substantially raise the typical phase-space
island width of the incoherent modes at which quasilinear
stabilization occurs.

Ion instability drive does not change the conclusion that incoherent
quasilinear flattening of the electron dimple would occur at
fluctuation levels that are too low to make significant non-linear
modification to the ion distribution. The linear drive of the combined
electron and ion distributions is brought to zero, if the ion
velocities are significantly higher than the ion-ion stability
threshold $c_s$, well before the quasilinear diffusivity of the ions
is significant, and before entrainment of the ions into typical mode
sizes. This conclusion is consistent with the code
observation\cite{Haakonsen} of a sustained initial period when the ion
distribution evolution is quiescent and laminar, and the electron
fluctuations are predominantly localized to their phase-space
separatrix. But it fails to explain the coherent structures that are
observed to grow and entrain the ions even well before the ion beam
velocities have slowed into the unstable regime.

\section{Electron Holes}
\label{holesect}

\subsection{Hole Structural Relationships}

At the other end of the spectrum of treatments of non-linear effects
and turbulence, far from the quasilinear diffusion approach, lies the
phenomenon of phase-space ``holes''. Conventionally such a
hole\cite{Gurevich1968,Berk1970,Schamel1986,Dupree1982a,Maslov1993,Goldman2003,Eliasson2006} refers to a localized coherent perturbation of the
distribution function phase-space density that is \emph{self-binding}
via its self-consistent potential. 
A perturbation to a one-dimensional
electron distribution function $f_e(y,v)$ can be self-binding if it
traps electrons. To do so it must give rise to an electric potential
that is positive in the hole. That requires the perturbation of the
phase-space density, denoted $\tilde f_e$ to be negative: a deficit of
electrons. (Likewise ion holes require a negative $\tilde f_i$.)
The parallel spatial coordinate is written $y$, for consistency with previous
sections.  Velocities $v$ are unnormalized in this section.

Electron holes in one dimension can exist at spatial scales from less than the
Debye length, upward.  The hole is essentially a
Bernstein-Greene-Kruskal (BGK) mode\cite{Bernstein1957}: a trapping
structure that self-consistently satisfies the Vlasov-Poisson system
of equations. There is substantial freedom in the form that such modes
can take. Entropy arguments\cite{Dupree1982a} support what is more
often proposed as an ansatz\cite{Schamel1979} that the velocity
dependence of $f_e$ in the trapped region is approximately parabolic
(with positive curvature, negative temperature) leading to what is
called a Maxwell-Boltzmann hole. They suggest that the most
probable spatial extent of a shallow electron hole is approximately 4 times
the plasma shielding length. However, the precise shape of the hole
proves not to have a major effect on its properties, and modeling the
hole as a rectangular box in $y,v$ space, of constant depth, yields
parameters that differ little from the Maxwell-Boltzmann hole\cite{Dupree1982a}. It is therefore
plausible to approximate the hole's shape with simple model functions,
and still expect to arrive at scalings that have reasonable
quantitative validity. Moreover, deep holes with spatial extent much larger
than the shielding length are possible.

The self-consistent Poisson's equation for an electron hole in one
dimension of space may be written
\begin{equation}
  \label{Poisson}
  \left(-{\partial^2\over\partial y^2} +{1\over
    \lambda^2}\right)\phi = {\rho\over\epsilon_0}
\end{equation}
where $\lambda$ is the plasma shielding length, and $\rho$ is the
charge-density of the hole. The shielding length would normally be
thought of as the Debye length $\lambda_{De}=\sqrt{\epsilon_0 T_e/n
  e^2}$ but it can be generalized to account for ion shielding and for
arbitrary distribution functions by regarding it as arising from the
medium's polarization term which is responsible for the dielectric
susceptibility\cite{Dupree1982a}. The real part of the linearized susceptibility for wave
number $k$ and phase velocity $\omega/k=v_p$ is then
\begin{equation}
  \label{shieldsuscept}
  \Re(\chi) = {1\over k^2\lambda^2} = \sum_{species}{\omega_p^2\over k^2} {\cal P}\int
  {df_b\over dv} {dv\over v_p - v},
\end{equation}
and this is the definition of $\lambda$.  $f_b$ denotes the
unperturbed background distribution away from the hole; ${\cal P}$ denotes the
principal value of the integral, and contributions like this from both
electrons and ions are included.

\begin{figure}[htp]
  \centering
  \includegraphics[width=0.5\textwidth]{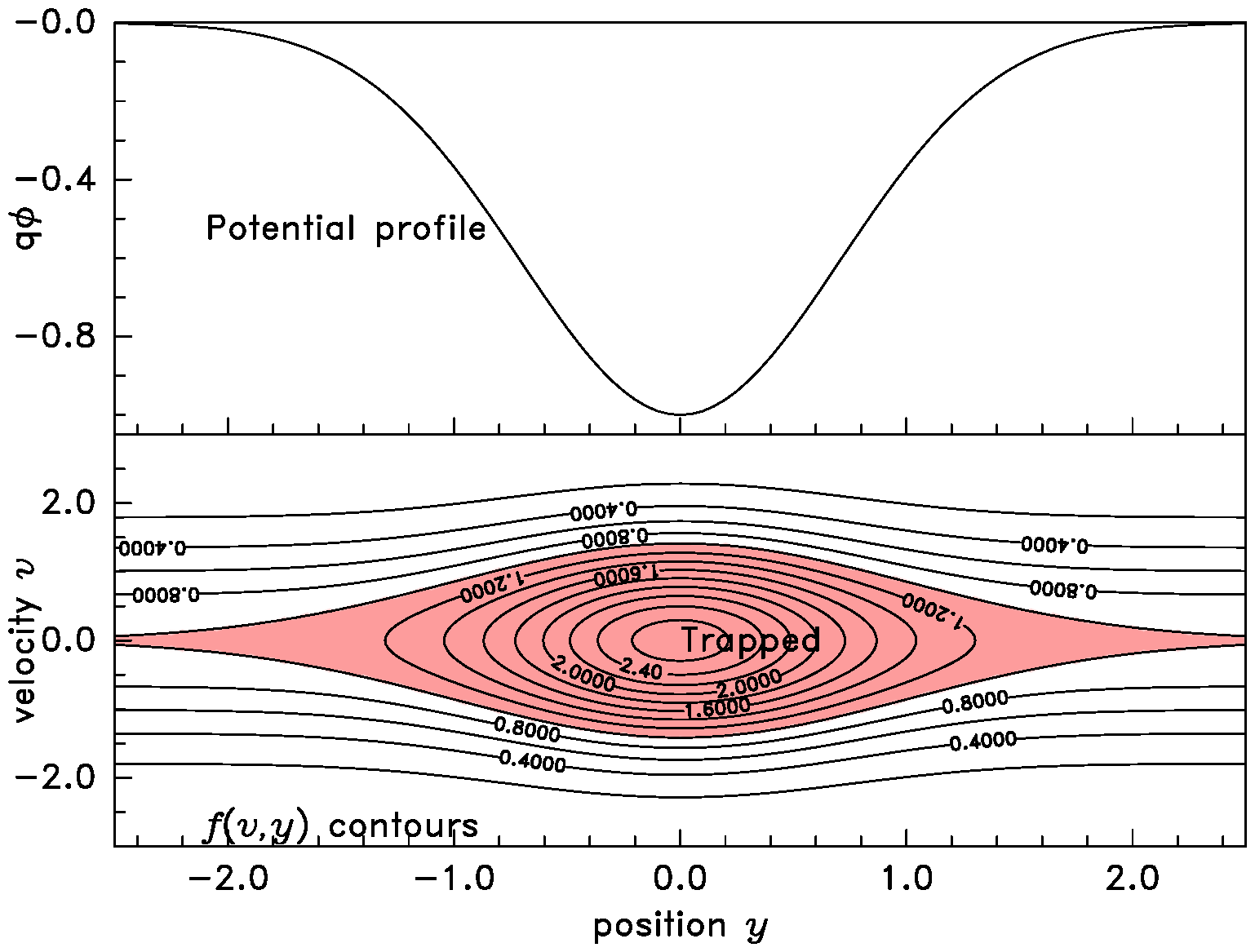}
  \caption{Contours of constant $f$ are contours of constant total
    energy in phase space. An attractive potential produces an island,
    which can become a hole. The parallel spatial extent of $y$ here
    should be considered far less than the width of the wake shown in
    Fig.\ \protect\ref{fig:hillcontour}.}
  \label{phaseisland}
\end{figure}

\begin{figure}[htp]
  \centering
  \includegraphics[width=0.45\hsize]{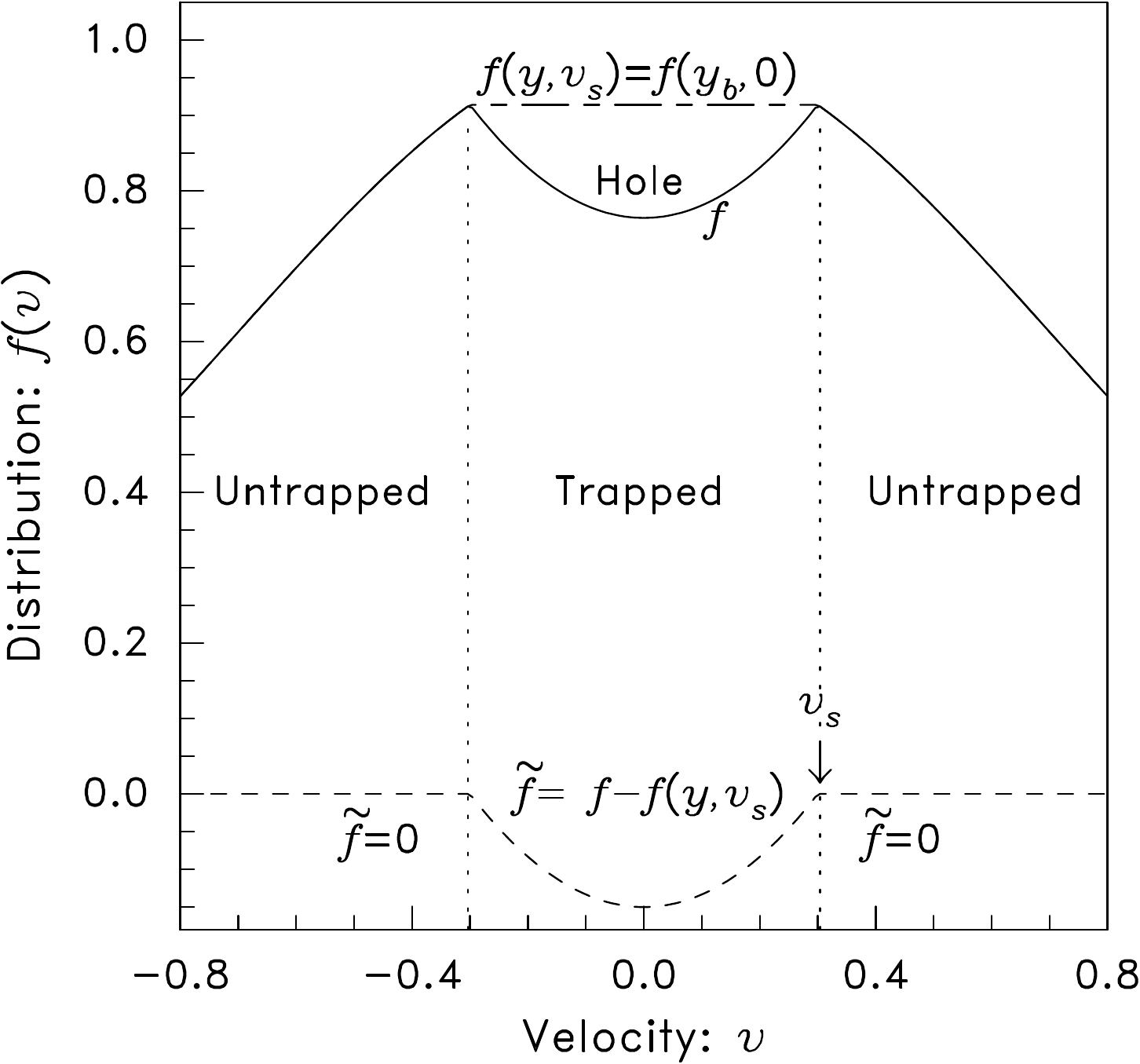}
  \caption{Schematic illustration of the distribution function $f(v)$
    in an electron hole. The reference distribution is uniform
    within the trapped region ($|v|<v_s$) and the difference $\tilde{f}$ is what
    determines the binding charge density.\label{distfig}}
\end{figure}

We consider a localized peaked hole potential structure that is the
solution of Poisson's equation (\ref{Poisson}). The potential energy
then has a well, which Fig.\ \ref{phaseisland} illustrates schematically.
For simplicity, and because it is the important case here, we
take the hole to be stationary (corresponding to $v_p=0$), though moving
structures can naturally be treated by a change of reference frame. It
gives rise to phase-space orbits of electrons (along which $f_e$ is
constant) that are the contours of
constant kinetic plus potential energy, so they satisfy
\begin{equation}
  \label{EnergyCons}
  {1\over 2}m_ev^2 + q \phi = const. = {1\over 2}m_ev_b^2
\end{equation}
where we write $q$ for the electron charge (it is negative), and $v_b$
is the velocity at a distant unperturbed (``background'') position ($y_b$) far from the
hole, where the potential is $\phi=0$ (so this potential is measured relative
to the background plasma potential in the vicinity of the
hole).  Since the
potential energy $q\phi$ is negative, orbits that connect to $y_b$
have a minimum speed at any position given by
\begin{equation}
  \label{minvel}
  v_s(y) = \sqrt{-2q\phi(y)\over m_e},
\end{equation}
which is the boundary of the trapped-electron island in phase space.
The charge density to be used in eq.\ (\ref{Poisson}) is then 
\begin{equation}
  \label{chargedensity}
  \rho(y) = q \int \tilde f(y,v) dv ,
\end{equation}
where $\tilde f$ is the change of $f$. It is zero ($\tilde f = f(y,v) -
f(y_b,v_b)=0$, $|v|>v_s$) on \emph{untrapped orbits}
, while on \emph{trapped
  orbits} $|v|<v_s$,
\begin{equation}
  \label{trappedpert}
  \tilde f(y,v) = f(y,v) - f(y,v_s) = f(y,v)-f(y_b,0).
\end{equation}
See Fig.\ \ref{distfig}.
The total electron density in the presence of the potential
perturbation would be different at position $y$ from its value at $y_b$ even if
$\tilde f$ were everywhere zero, because $\int f(y,v) dv\ne \int
f(y_b,v)dv$. However, in Poisson's equation (\ref{Poisson}), that difference is
contained in the $\phi/\lambda^2$ term, not in $\rho$. It is the
linearized dielectric response of the plasma. In eq.\ (\ref{Poisson})
the $\rho$ contains only the charge density attributable directly to
$\tilde f$.

Now we introduce a lumped-parameter model of the hole in which $\tilde
f$ has a characteristic magnitude at the center of the hole $\tilde
f_0=\tilde f(0,0)$, and the characteristic widths in velocity, $v$,
and space, $y$, of the trapped region are $\Delta v$ and $\Delta y$
respectively.  We define $\Delta v$ so that the charge density at the
hole's spatial center ($y=0$) is
\begin{equation}
  \label{rhofromf}
  \rho=\rho(0) = q \tilde f_0 \Delta v.
\end{equation}
In Poisson's equation the term $\partial^2\phi/\partial y^2$ will be
of magnitude approximately $\phi(0)/\Delta y^2$. But if the hole has
large spatial extent, $\Delta y\gg
\lambda$, that term is negligible and we find as a \emph{quasi-neutral}
approximation to the hole
\begin{equation}
  \label{Poisapprox}
  {\phi\over \lambda^2} \approx {\rho\over\epsilon_0} =
  {q\over \epsilon_0}\tilde f_0 \Delta v.
\end{equation}
(The quasi-neutral hole can be considered to be two ``double-layers''
that trap electrons between them. A small hole that is not
quasi-neutral can be analysed\cite{Dupree1982a,Schamel1986} to find a
comparable relationship between $\phi$ and $\Delta v$, in which the
proportionality coefficient depends upon $\Delta y/\lambda$. So our
conclusions are not qualitatively changed for holes of small $\Delta y$.)

Hereafter, we refer to values at $y=0$ and we drop the
repetition of this fact in our notation.
There is a proportionality between the two measures of the hole
velocity width $v_s$ and $\Delta v$. It requires knowledge of the
velocity-shape of the hole to obtain its exact coefficient. For example, if $\tilde f$
is uniform throughout the trapped region, then $ \Delta v = 2 v_s$
while if $\tilde f$ is parabolic in the trapped region, then $\Delta v
= {4\over 3} v_s$, and if it is triangular $\Delta v = v_s$. Adopting
this last alternative, for reasons that will become clear later, we have
\begin{equation}
  \label{vsize}
\Delta v = v_s = \sqrt{-2q\phi\over m_e},
\end{equation}
and the relationship between the depth and
velocity-width of the hole becomes
\begin{equation}
  \label{depthwidth}
 - \tilde f = {\epsilon_0 m_e\Delta v\over 2q^2\lambda^2} = {n_b \Delta
    v\over 2 \omega_{pe}^2\lambda^2} = 
  \left(\lambda_{De}\over \lambda\right)^2 {n_b \Delta v\over
    v_{te}^2}
  = { \sqrt{\pi}}  \left(\lambda_{De}\over \lambda\right)^2
  f_b(0) {\Delta v\over v_{te}}.
\end{equation}
where $v_{te}\equiv \sqrt{2T_e/m_e}$ and parameters such as $\lambda_{De}$,
$\omega_{pe}$, $f_b$ and $n_b$
refer to the background plasma. Different assumptions about hole
profile shape will somewhat change the coefficient. But in
general, when the shielding length is not too different from the Debye
length, the fractional hole depth $\tilde f / f_b$ is roughly
equal to the fractional velocity-width $\Delta v/v_{te}$ for a
quasi-neutral ($\Delta y\gg \lambda$) hole.

Since $f$ cannot be negative there is a maximum hole depth and size
obtained by setting the trapped distribution function (phase-space
density) equal to zero i.e.\ $-\tilde f =f_b(0)$. Then
\begin{equation}
  \label{finalwidth}
  {\Delta v\over v_{te}} = {1\over \sqrt{\pi}} \left(\lambda\over
    \lambda_{De}\right)^2 .
\end{equation}

\subsection{Hole Growth}

In view of the proportionality between the hole depth $|\tilde f|$ and
its velocity-width $v_s$, for a hole to grow in velocity width (and
hence in potential) it must become deeper. When this happens in an
effectively collisionless plasma, the shape of the hole, $\tilde
f(v)$, is determined not by maximizing entropy but by the constancy of
$f$ on orbits. The absolute value of $f$ on trapped orbits ($f_t$) is
an invariant function\cite{Gurevich1968} of the orbit's action ($\int v dx$),
provided the hole phase-space orbits remain closed. (Fine-scale mixing
does not substantially change the mean $f$ on a phase-space orbit, and
so does not escape this constraint. In recognizing it we abandon
decisively the common presumption that the hole remains parabolic.) Therefore
the only way for a hole to become deeper is that the \emph{external}
distribution function $f(y,v_s)= f(y_b,0)$ \emph{increases}. In
Dupree's analysis\cite{Dupree1983} of growth of moving ($v_p\not=0$)
holes in an electron-ion instability, the way the external $f$
increases is by the hole decelerating to lower $|v_p|$ so that
$f(y_b,v_p)$ increases (e.g. for a Maxwellian external distribution).
In the present context, however, there is a different mechanism
inducing hole growth. It is that the plasma is drifting in the $x$
direction, perpendicular to the magnetic field. Consequently there is
a convective time derivative of the external density: $v_x{\partial
  f\over \partial x}$; and this is indeed positive in the wake. In
other words, the drift de-energization term in the Vlasov equation
that has the effect of generating the dimple continues to operate if a
hole is present, and is a cause of hole-growth. Or equivalently,
perhaps conceptually simpler still, the hole experiences a background
plasma of rising density because of drift. The importance of these
remarks is that the growth of a hole is not suppressed quasilinearly
by reaching sufficiently strong perturbation that the distribution
function is flattened. The hole is coherent; and as long as that
coherence is maintained, it continues to grow as $f(y_b,0)$
grows\footnote{$y_b$ here should be considered to be a distance large
  compared with the spatial extent of the hole but small compared with
  the width of the wake.}.

The only way a hole stops growing short of maximal size, assuming
there is insufficient turbulence to tear it apart, is for it to
convect out of the spatial region where the external $f$ is
growing. Holes move mostly along the direction of the (unperturbed)
phase space orbits. Therefore their parallel ($y$-) motion leaves them
at approximately constant $f$. When a hole moves away from the peak of
the wake's potential profile, it is therefore swept out of the wake,
at approximately the electron parallel velocity, without any
consequent growth. Once the hole reaches the unperturbed plasma
outside the wake, no convective growth term is operating, and it will
move away without further growth.

Therefore, once a hole has formed with sufficient coherent integrity,
the only condition for it to grow is that it stays inside the wake,
which in general means it must remain near the peak of the wake's
potential energy curve (bottom of its electrostatic potential well)
$y=0$.  That hole position is unstable, so most holes will move from
it, and then be convected out of the wake before they've grown large,
because the electron orbit duration in the $x$-direction is rather
small except when they are on axis. But a few may remain at the wake
axis long enough to grow to near maximal size. It is those we now analyse.

There is a linear relationship between $\tilde f$ and $v_s$ if
$\lambda$ and the hole shape can be approximated as constant. Since
the growing edge of a hole entrains additional phase-space area on
which the distribution is equal to $f_b(0)$, the hole velocity profile
in this approximation is triangular. That was the basis for choosing
the triangular profile in the previous section.  But we now do a
self-consistent calculation that shows what the shape in velocity
space of a growing quasi-neutral hole profile actually is. This
requires a treatment that is self-consistent and, for a deep hole,
non-linear (i.e.\ avoiding the commonly used linearized plasma response
explained in the prior section). It is most simply performed for a
quasi-neutral hole by setting the net charge density to zero as
follows.

Consider a background plasma with Maxwellian electrons,
\begin{equation}
  \label{Maxwellian}
  f_b(v) = {n_b\over v_{te}\sqrt{\pi}} \exp(-u^2),
\end{equation}
where $v_{te}\equiv 2T_e/m_e$ and $u\equiv v/v_{te}$. Write the normalized
phase-space separatrix speed $u_s$ at potential $\phi$:
$u_s^2=-q\phi/T_e$. This is positive because $\phi$ is positive and
$q$ negative for electrons. Take the reference flat-top electron distribution
to be constant within the trapped region, and constant along untrapped
orbits as shown in Fig.\ \ref{distfig}:
\begin{equation}
  \label{Flattop}
    f_f(v) =  {n_b\over v_{te}\sqrt{\pi}}\times\left\{
  \begin{array}{ll}
    \exp(u_s^2-u^2)&\mbox{for $u\ge u_s$}\\
    1 &\mbox{for $u<u_s$}\\
   \end{array}\right. .
\end{equation}
This is the distribution that would arise if an electron-trapping
potential hill arose slowly (compared with the electron bounce time)
in a background distribution that was \emph{not} varying with time.
The density of this distribution is a function of the normalized
potential, $u_s^2$. It can readily be evaluated\cite{Gurevich1968} as
\begin{equation}
  \label{electrondens}
  n_f(u_s^2) =\int f_f dv= n_b\left[{2u_s\over \sqrt{\pi}}+{\rm
      e}^{u_s^2}\erfc(u_s) \right].
\end{equation}

The ions in the wake can be represented quite
well\cite{Gurevich1969,Haakonsen} by two ion streams, each of narrow spread
in speed. Near the wake axis they are equal and opposite. We take
their Mach number outside the hole to be $M=v_b/c_{s}$. Inside the
hole the ion speeds are lower, because the hole repels ions, and by
conservation of energy the mach number there is
$\sqrt{M^2-2u_s^2}$. Hence, by conservation of flux, the ion density
is
\begin{equation}
  \label{iondens}
  n_i(u_s^2) = n_b {M \over \sqrt{M^2-2u_s^2}}
\end{equation}

If the actual electron distribution in the trapped region is $f_t(v)$,
different from the reference $f_f$ by $\tilde{f}$, then
quasi-neutrality can be expressed as the cancellation of the charge
arising from the background density in the perturbed potential,
$n_f-n_i$, and the hole charge density $\propto\int_0^{u_s}
\tilde{f}du$. That is,
\begin{equation}
  \label{holequasi}
  n_f-n_i=n_b H(u_s)/\sqrt{\pi} = v_{te} f_{b} H(u_s) = 2 v_{te}
  \int_0^{u_s}f_{b}-f_t \;du,  
\end{equation}
where in the trapped-region $f_f=f_b$ is independent of $u$; $f_b$
refers to the background value $f_b(0)$ at $u=0$; and the function
$H$, for any given $M$, is essentially the normalized reference
charge-density from both electron and ion distributions,
\begin{equation}
  \label{Hfunction}
  H(u_s)= \sqrt{\pi}\left[{2u_s\over \sqrt{\pi}}+{\rm
      e}^{u_s^2}\erfc(u_s) - {M \over \sqrt{M^2-2u_s^2}}\right],
\end{equation}
which must be cancelled by the hole charge-density.
\begin{figure}[htp]
  \centering
  \includegraphics[width=0.5\textwidth]{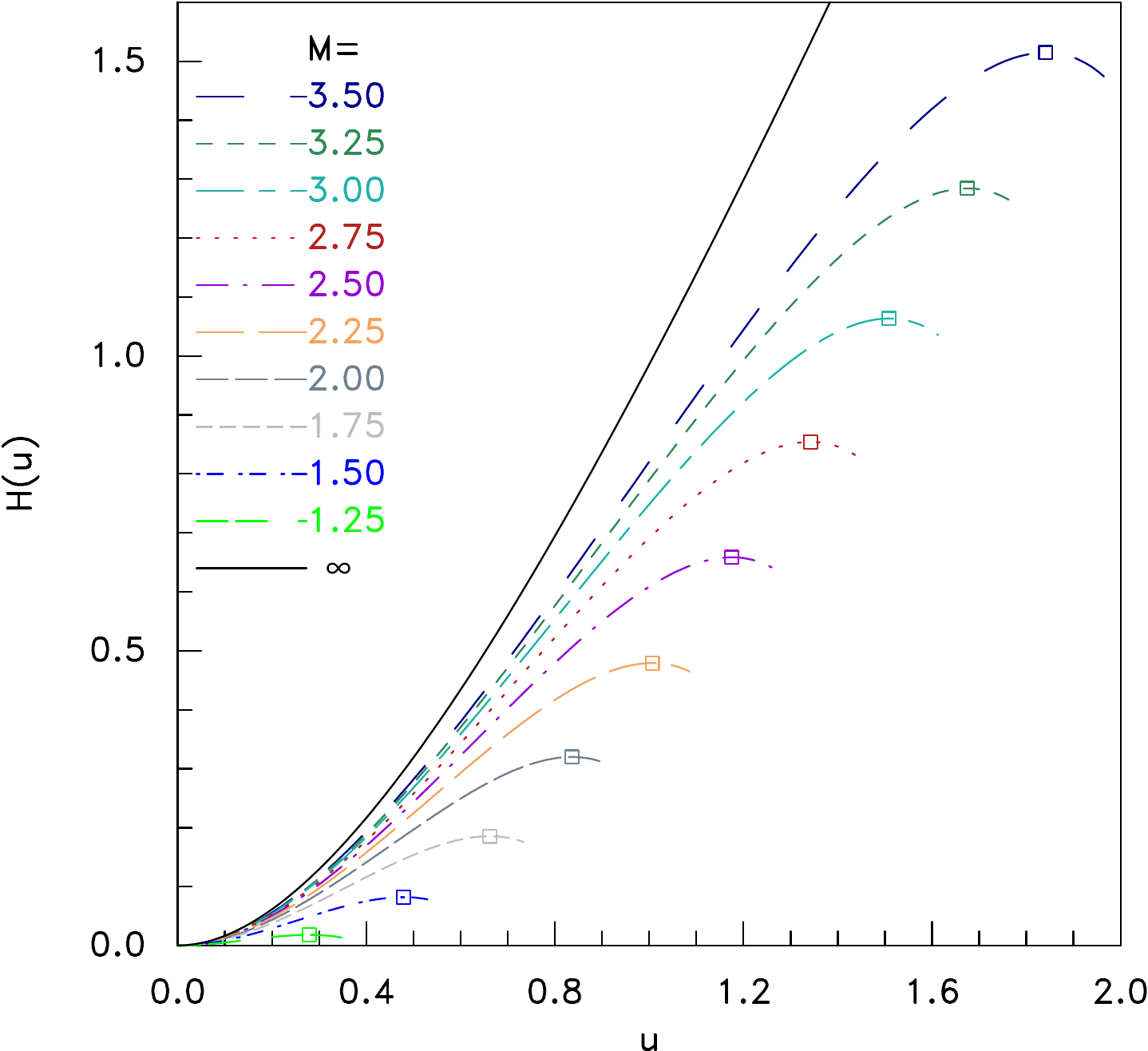}
  \caption{The charge-density function $H(u)$ for various values of
    ion mach number $M$. The solid line marked $\infty$ corresponds to
    an immobile ion background density independent of potential.}
  \label{Hfigure}
\end{figure}
Figure \ref{Hfigure} shows the form of $H(u)$.

Equation (\ref{holequasi}) determines the relationship between the hole
velocity-width, $u_s$, and the changing background electron density expressed
as the peak of its Maxwellian, $f_{b}$. In the context of a growing
hole, the actual trapped electron distribution $f_t$ remains
invariant once formed. Therefore we can differentiate the equation
with respect to the hole width, $u_s$, and it becomes
\begin{equation}
  \label{holediff}
  {d\over du_s}[H(u_s)f_{b}] = 2 u_s{d\over du_s} f_{b}.
\end{equation}
No contribution arises to the differential from the fixed
$f_t(u)$, and none comes from the limit because $f_{b}=f_t$ at
$u=u_s$: the newly trapped electrons have phase-space density equal to the
instantaneous background density. This equation can be written as a
simple quadrature
\begin{equation}
  \label{quadrature}
  \int {df\over f} = \int {dH\over 2u-H}.
\end{equation}
The derivative $H'(u)= dH/du$ is positive at moderate $u$, but reaches zero at a
certain value $u=u_{max}$ dependent on $M$. This is where
the hole reaches its maximum possible size. At that size the rate of
hole width-increase with respect to background density becomes
infinite: the hole blows up. Its growth can no
longer be described by the quasi-neutral hole equilibrium equations.
We call the corresponding
value of the background distribution $f_{b}=f_{max}$. And we find a
set (at various $M$) of universal curves by numerical integration
from $u_{max}$ backwards toward zero. So that
\begin{equation}
  \label{integrated}
  \ln(f/f_{max}) = \int_{u_{max}}^u {H'\over 2u-H} du.
\end{equation}
\begin{figure}[htp]
  \centering
  \includegraphics[width=0.45\textwidth]{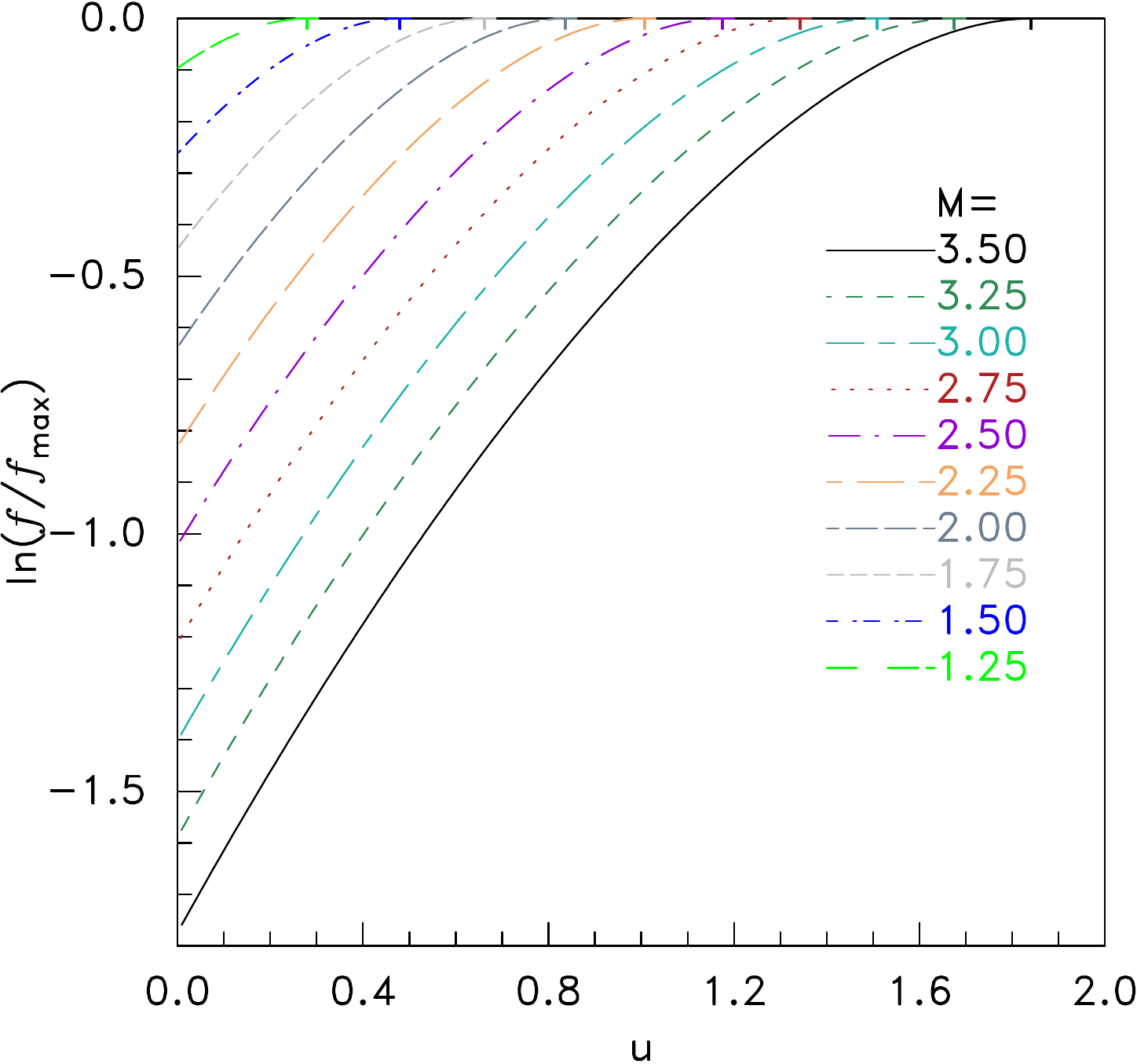}
  \caption{Trapped electron distribution function forms for various
    ion Mach numbers. These also give the hole growth curves arising
    in growing background density.}
  \label{trappeddist}
\end{figure}
Fig.\ \ref{trappeddist} shows the result. The significance of a curve
for some given $M$ is this. As a hole moves from one value of $f$ to a
larger value, the hole size, $u_s$, grows by moving along the
curve. As it does so, the trapped distribution function is built up
within the hole by incrementally trapping additional phase-space. The
shape of $f_t(u)$ is therefore also given precisely by this functional
dependence. So these curves can be considered to represent the shape in
velocity space ($u$) of a hole that grows from infinitesimal size up
to any finite current size $u_s$. The boundary of the hole is $u_s$,
at which $f_t=f_{b}$, and inside the hole $f_t(u)= f_{b}
\exp[\ln(f(u)/f_{max})-\ln(f_{b}/f_{max})]$. If the hole started at
finite size $u_{init}$, with a trapped distribution inside $u_{init}$
different from the growing hole form, nothing changes except in the
initial region of the hole $u\le u_{init}$. It must begin from a trapped distribution
that satisfies the quasi-neutrality equation (\ref{holequasi}) but
its initial shape will be determined by whatever mechanisms governed
its formation. Thereafter, as it grows incrementally, driven by rising
external density, it follows the curve, and the trapped distribution
function is built up accordingly.

No hole can grow stably beyond $u_{max}$, the place where the curves'
gradient becomes zero. That size is the maximum stable hole size. When
it is reached, the hole blows up, and disruption of the ion streams
will take place: their large free energy will be released through
additional non-linear processes not described here. It is found
that an excellent fit to the numerical values of $u_{max}$ for the
range of $M$ shown is
\begin{equation}
  \label{umaxeq}
  u_{max} = M/\sqrt{2}-0.6.
\end{equation}

A hole that begins at a position with a certain value of $f_{b}$ or
equivalently background density $n_b$, will grow as $n_b$ grows,
provided it remains near the wake potential ridge, not convecting out
of the wake.  It will reach the disruptive size when the density has
increased by a factor that can be read off the curves of Fig.\
\ref{trappeddist}. For example, the $M=2.5$ curve has
$\ln(f/f_{max})=-1$ at $u=0$, so a small hole will reach disruptive
size when the density has increased by a factor $\exp(1)$. Or again
for $M=2$, $\ln(f/f_{max})=-0.5$ at $u=0.1$, so a hole that starts with
size $u_s=0.1$ will grow to disruption when $n/n_{init}=\exp(0.5)$.

The wake axis experiences a large increase of background density
($n_b=n_\infty \exp(e\phi/T_e)$) as its potential subsides from the
large negative values immediately behind the object. It is clear,
therefore, that holes formed in this region that remain at the axis
have sufficient density increase to reach disruptive size. This
conclusion contrasts with the demonstration that quasilinear diffusion
cannot reach a level of strong ion perturbation.

These analytic conclusions are in accord with the observations in the
simulations. Small electron holes form in the dimple (by mechanisms we
don't here calculate in detail). While they remain near the axis and
are not convected out of the wake, they grow. Some eventually grow
large enough to disrupt the ion streams.

\section{Summary}

The wake behind an object in a magnetized plasma with predominantly
cross-field drift, and short Debye length, experiences one-dimensional
electrostatic instabilities. The electron velocity distribution along
the field acquires a depression we call the dimple, on orbits that
spend a long time near the axial ridge of the potential energy
structure of the wake. The driving term of this unstable dimple can
equivalently be regarded as either de-energization of the electrons by
drift perpendicular to $B$, down the potential energy ridge; or, as
drift in an increasing background density, filled in by parallel
velocity less quickly for orbits with low parallel velocity near the
ridge. The term may be approximated in the ``Krook'' collisional form
in Boltzmann's equation.

The second viewpoint, of drift into increasing background density,
provides a more transparent understanding of the resulting non-linear
dynamics. The collisionless form of the dimple is immediately unstable
to electrostatic waves near the wake axis, having phase velocities
lying within the dimple velocity width of approximately
$(m_e/m_i)^{-1/4}c_s=(m_e/m_i)^{1/4}v_{te}/\sqrt{2}$ (approximately $7c_s$ for
hydrogen plasmas). The waves will grow and, if incoherent, will fill
in the electron distribution function dimple until it becomes
marginally stable.  The ion parallel velocity distribution consists of
two streams attracted inward toward the wake axis. They contain a
great deal of free energy. But, because the stream velocity spacing is
large close to the object, the ion distribution is not itself linearly
unstable until far downstream. Nevertheless, it is observed in
numerical simulations (and sometimes in space) that large-amplitude
perturbations grow and substantially disrupt the ion streams, long
before they have become linearly unstable.

The electric fields associated with quasilinear velocity-space
diffusivity sufficient to stabilize the electron distribution dimple
are too small to cause substantial perturbation of the ions. The ion
disruption therefore cannot be explained by linear or quasilinear
instability growth. However, the driving term of electron instability
is effective also in causing the non-linear growth of electron holes.
Holes formed away from the wake axis, or that move away from it, leave
the wake at approximately the electron phase-space separatrix orbital
speed before they can grow very much.  Some holes, however, are formed
at the wake axis and remain near it. When they do, the continuing
background density enhancement causes them to grow to a maximum size
beyond which they explode and disrupt the ion streams. We have
calculated the form, the growth, and the maximum size of such electron
holes, establishing a (to our knowledge) new theoretical non-linear instability
mechanism associated with cross-field drift into higher density
regions. Electron holes grown by this mechanism provide the missing
piece of the wake stability puzzle. Coherent hole growth is not
suppressed quasilinearly, and explains how large, ion-disrupting,
perturbations can occur before the ion streams become themselves
linearly unstable.

Undoubtedly, there are many other important phenomena in the moon
wake, including Alfv\'enic processes that the present electrostatic
treatment omits. But, because of their large linear growth rate, we consider
the \emph{electrostatic} instabilities to be primary.  Several important
details of the explanation remain to be investigated. We have not
addressed the question of exactly how small electron holes form in the
first place, nor have we quantitatively analysed their positional
stability, which decides whether or not they remain at the wake axis
and grow. Moreover, the present treatment, limited to parallel
one-dimensional dynamics, omits oblique wave-vector perturbations,
which might in some circumstances be important.  Nevertheless, the
qualitative agreement with phenomena observed in one-dimensional
simulations provide strong evidence that the electron hole growth
mechanism is the key explanation of these simulations at least.

\section*{Acknowledgements}

This work was partially supported by the NSF/DOE Basic Plasma Science
Partnership under grant DE-SC0010491.

\bibliography{mybib,Haak}

\end{document}